\documentstyle[a4,12pt,epsf]{article}

\begin{document}

\newcommand{\elf}{ELFE}
\newcommand{\beq}{\begin{equation}}
\newcommand{\eeq}{\end{equation}}
\newcommand{\beqa}{\begin{eqnarray}}
\newcommand{\eeqa}{\end{eqnarray}}
\newcommand{\half}{\frac{1}{2}}
\newcommand{\gsim}{\buildrel > \over {_\sim}}
\newcommand{\lsim}{\buildrel < \over {_\sim}}
\newcommand{\ie}{{\it ie}}
\newcommand{\eg}{{\it eg}}
\newcommand{\cf}{{\it cf}}
\newcommand{\etal}{{\it et al.}}
\newcommand{\gev}{{\rm GeV}}
\newcommand{\jpsi}{J/\psi}
\newcommand{\order}[1]{${\cal O}(#1)$}
\newcommand{\morder}[1]{{\cal O}(#1)}
\newcommand{\eq}[1]{Eq.\ (\ref{#1})}
\newcommand{\ptr}{p_T}
\newcommand{\as}{\alpha_s}
\newcommand{\ket}[1]{\vert{#1}\rangle}
\newcommand{\bra}[1]{\langle{#1}\vert}
\newcommand{\qpair}{q\bar q}

\newcommand{\PL}[3]{Phys.\ Lett.\ {{\bf#1}} ({#2}) {#3}}
\newcommand{\NP}[3]{Nucl.\ Phys.\ {{\bf#1}} ({#2}) {#3}}
\newcommand{\PR}[3]{Phys.\ Rev.\ {{\bf#1}} ({#2}) {#3}}
\newcommand{\PRL}[3]{Phys.\ Rev.\ Lett.\ {{\bf#1}} ({#2}) {#3}}
\newcommand{\ZP}[3]{Z. Phys.\ {{\bf#1}} ({#2}) {#3}}
\newcommand{\PRe}[3]{Phys.\ Rep.\ {{\bf#1}} ({#2}) {#3}}

\begin{titlepage}
\begin{flushright}
        NORDITA--96/76 P\\
        hep-ph/9611436\\
       Revised December 19, 1996
\end{flushright}

\vskip 2.5cm

\centerline{\Large \bf Physics Opportunities at ELFE\footnote{Talk given at
the meeting on {\em Future Electron Accelerators and Free Electron Lasers},
Uppsala, April 25-26, 1996. Work supported in part by the EU/TMR contract ERB
FMRX-CT96-0008.}}

\vskip 1.5cm

\centerline{\bf Paul Hoyer}
\centerline{\sl Nordita}
\centerline{\sl Blegdamsvej 17, DK--2100 Copenhagen \O, Denmark}

\vskip 2cm

\begin{abstract}
I review some central physics opportunities at the $15\ldots 30$ GeV
continuous beam electron accelerator \elf, proposed to be built in conjunction
with the DESY linear collider. Our present detailed knowledge of
single parton distributions in hadrons and nuclei needs to be supplemented
by measurements of compact valence quark configurations, accessible through
hard exclusive scattering, and of compact multiparton subsystems which
contribute to semi-inclusive processes. Cumulative ($x>1,\ x_F>1$) processes
in nuclei measure short-range correlations between partons belonging to
different nucleons in the same nucleus. The same configurations may
give rise to subthreshold production of light hadrons and charm.
\end{abstract}

\end{titlepage}

\newpage
\renewcommand{\thefootnote}{\arabic{footnote}}
\setcounter{footnote}{0}
\setcounter{page}{1}

\section{Introduction}

The ELFE@DESY project aims at utilizing a future DESY linear electron
collider~\cite{wiik} to accelerate electrons to $15\ldots 30$ GeV and then use
the HERA electron ring to stretch the collider bunches into an intense
(30 $\mu$A) continuous extracted beam~\cite{frois}. Polarized electrons
will be scattered from both light and heavy fixed targets, with
luminosities in the ${\cal L}= 10^{35} \ldots 10^{38}$ cm$^{-2}$s$^{-1}$
range. In this talk I discuss some of the central physics issues that
can be addressed with this type of accelerator. Since \elf\ experiments are
many years in the future I shall concentrate on questions related to basic
aspects of QCD, which will remain of fundamental interest and which require
the capabilities of an accelerator like
\elf.

\begin{table}[hb]{\centerline{{\rule[-3mm]{0mm}{8mm}}
Table 1. {\bf Features and opportunities of an \elf\ accelerator.}}}\bigskip
\begin{tabular}{p{2cm}llll}{\rule[-3mm]{0mm}{8mm}}
&{\bf Features} & {\bf Opportunities} &\\
\cline{2 - 3} {\rule[-3mm]{0mm}{8mm}}
&High luminosity & Study rare configurations &\\
&$\cal{L}\sim$ $10^{35} \ldots 10^{38}$ cm$^{-2}$s$^{-1}$ &\ of target wave
function &\\
\cline{2 - 3} {\rule[-3mm]{0mm}{8mm}}
&Energy & Perturbative QCD &\\
&$E=15 \ldots 30$ GeV & Resolution of \order{0.1\ {\rm fm}} &\\ 
& & Charm production &\\
\cline{2 - 3} {\rule[-3mm]{0mm}{8mm}}
&High duty factor $\sim 80\%$ & Event reconstruction &\\
\cline{2 - 3} {\rule[-3mm]{0mm}{8mm}}
&High energy resolution & Exclusive reactions &\\
&$\Delta E/E \simeq 5 \cdot 10^{-4}$ & Inclusive reactions at high $x$ &\\
\cline{2 - 3} {\rule[-3mm]{0mm}{8mm}}
&Polarization & Amplitude reconstruction &\\
& & Spin systematics of QCD &\\
\cline{2 - 3} {\rule[-3mm]{0mm}{8mm}}
\end{tabular}
\end{table}

In Table 1 I list the main features of the \elf\ accelerator, and the
opportunities that they provide. Compared to existing electron and muon beams,
the advantages of \elf\ are in luminosity (compared to the muon beams at CERN and
Fermilab), in duty factor (compared to SLAC) and in energy (compared to TJNAF).
Competitive \elf\ experiments will rely on a combination of these strong features.
The HERMES experiment at DESY works in the same energy range but at a lower
luminosity and duty factor compared to ELFE. HERMES will prepare the ground
for \elf\ physics, together with experiments at TJNAF in the U.S., GRAAL in
Grenoble and the lower energy electron facilities ELSA (Bonn) and MAMI (Mainz).

As I shall discuss below, an important part of physics at \elf\ will deal with
exclusive reactions, or with inclusive reactions at large values of Bjorken
$x=Q^2/2m\nu$. The energy range of $15\ldots 30$ GeV is actually optimal for
such studies, as seen from the following argument. The inclusive deep inelastic
cross-section scales (up to logarithmic terms) in the virtuality $Q^2$
and energy $\nu$ of the photon like  
\beq
\frac{d^2\sigma_{DIS}}{dQ^2dx} \propto \frac{1}{Q^4}F(x) \label{siginc}
\eeq
Exclusive processes are still more strongly suppressed at large $Q^2$,
\eg,
\beq
\frac{d\sigma}{dQ^2}(ep\to ep) \propto \frac{F_p^2(Q^2)}{Q^{4}} \propto
\frac{1}{Q^{12}}\ \ \ \ \  (x=1) \label{sigexc}
\eeq
Typically we want to
reach at least $Q^2=\morder{10\ \gev^2}$ to be able to use perturbative QCD
(PQCD) and to have a resolution of \order{0.1\ {\rm fm}}. This implies
$\nu=\morder{5\ \gev}$ at large $x\simeq 1$. At \elf, such energies correspond
to the photon taking a moderate fraction $y=\nu/E_e \simeq 0.15 \ldots 0.3$ of
the electron energy, which is practical for measurements. This may be
contrasted with the situation at HERA, which is equivalent to a fixed target
experiment with an electron energy $E_e\simeq 50000$ GeV. A photon with energy
$\nu=5$ GeV would at HERA correspond to $y \simeq 0.0001$. It is clearly very
difficult to measure the large $x$, moderate $Q^2$ region at HERA, but it is
the natural territory of an accelerator in the \elf\ energy range.

In the following I shall discuss three aspects of physics at \elf\ which
relate to basic issues in QCD:
\begin{itemize} 
\item[$\bullet$] {\em Wave function measurements.} Most of our present
knowledge of hadron and nuclear wave functions stems from
hard inclusive scattering, which measures single parton
distributions. The phenomenology of hard exclusive scattering,
which is sensitive to compact valence quark configurations, is still in its
infancy. Although considerable progress may be expected in this field in
the coming years, the measurements are so demanding that an accelerator with
\elf's capabilities is sorely needed. On the theoretical front, we still do
not have a full understanding of which properties of the wave function are in
principle measurable in hard scattering. It seems plausible that
semi-inclusive processes can be used to  measure configurations where a
subset of partons are in a compact configuration, while the others are summed
over.

\item[$\bullet$] {\em Short range correlations in nuclei.} Scattering which is
kinematically forbidden for free nucleon targets has been experimentally
observed, and includes DIS at $x>1$, hadron production at Feynman $x_F>1$ and
subthreshold production processes. Such scattering requires short range
correlations between partons in more than one nucleon, and thus gives
information about unusual, highly excited nuclear configurations.

\item[$\bullet$] {\em Charm production near threshold.}\footnote{Due to space
limitations, this topic is not included in these proceedings, but will be
published separately \cite{charm}.} Production close to
threshold requires efficient use of the target energy and hence favors
compact target configurations. Heavy quarks are created in a restricted
region of space-time, where perturbative calculations are reliable. Both
features conspire to make the production of charm near threshold a sensitive
measure of new physics, including unusual target configurations and higher
twist contributions. The \elf\ accelerator will work in the region of charm
threshold $(E_\gamma \simeq 9\ \gev)$ and provide detailed information about
both charmonium and open charm production.
\end{itemize}

The above selection of physics topics is obviously far from complete. I refer
to earlier presentations of \elf\ physics \cite{elphys} as well as to the
review by Brodsky \cite{bro} for further discussions of these and other
aspects of QCD phenomenology. In particular, I shall not cover here the
important and topical area of color transparency, but refer to recent
reviews \cite{ct} and references therein.

\section{Wave function measurements}

\subsection{Inclusive Deep Inelastic Scattering}

Our most precise knowledge of nucleon (and nuclear) structure is based on
deep inelastic lepton scattering (DIS), $\ell N \to \ell' X$, and related hard
inclusive reactions. As is well-known, DIS measures the product of a
parton-level subprocess cross-section $\hat \sigma$ and a target structure
function $F$. Thus, schematically and at lowest order in the strong coupling
$\as$,
\beq
\frac{d^2\sigma(eN\to eX)}{dQ^2dx} = \hat\sigma(eq\to eq)\, F_{q/N}(x,Q^2)\,
[1+\morder{\as}]  \label{epex}
\eeq
The structure functions $F_{q/N}$ have been measured over an impressive range
in $x$ and $Q^2$, covering $.0001 \lsim x \lsim 1$ and $1 \lsim Q^2 \lsim
10000\ \gev^2$. Their logarithmic $Q^2$-dependence (`scaling violations')
predicted by QCD has been tested, and their `universality' verified, \ie, the
same structure functions describe other hard inclusive reactions such as $pp
\to jet+X$, $\pi^-p \to \mu^+\mu^- + X$, $pp \to \gamma + X$, etc. The many
detailed measurements and successful cross-checks have together established
QCD as the correct theory of the strong interactions, and made us confident
that basic properties of hadron wave functions can be deduced from experimental
measurements using the methods of PQCD.

The success of DIS phenomenology should not make us forget that the
structure function $F_{q/N}(x,Q^2)$, no matter how completely known, still
only provides us with a very limited knowledge of the nucleon wave function. In
terms of a (light-cone) Fock state expansion of the proton wave function,
\beqa
\ket{p}&=& \int\prod_i\, dx_i\, d^2k_{\perp i} \left\{
\Psi_{uud}(x_i,k_{\perp i}) \ket{uud} \right. \nonumber \\
&+& \left. \Psi_{uudg}(\ldots) \ket{uudg}+ \ldots + \Psi_{\cdots}(\ldots)
\ket{uudq\bar q}+ \ldots \right\} \label{fock}
\eeqa
the structure function $F_{q/p}$ can be expressed as a sum over the absolute
squares of all Fock components $n$ that contain a parton $q$ with the measured
momentum fraction $x$,
\beq
F_{q/p}(x,Q^2)= \sum_n \int^{k_{\perp}^2<Q^2} \prod_i\, dx_i\, d^2k_{\perp i}
|\Psi_n(x_i,k_{\perp i})|^2 \delta(x-x_q)  \label{strfn}
\eeq
Due to the average over Fock states, the most probable states will typically
dominate in the structure function. Information about partons which do not
participate in the hard scattering is lost in the sum of \eq{strfn}. The
structure function is a single parton inclusive probability distribution that
does not teach us about parton correlations. However, at large values of
$x$ the structure function singles out unusual Fock states where one parton
carries nearly all momentum, and all other partons therefore must have low $x$.

\subsection{Hard Exclusive Scattering}

Clearly, it is desirable to make also other measurements of hadron wave
functions. This is not as easy as it sounds, given that we only master the
perturbative region of QCD. We need to study a hard scattering, where the
subprocess can be identified and calculated, and where the dependence on the
soft wave function factorizes. The factorization between hard and soft
processes is a nontrivial feature in a theory like QCD with massless
(long-range) gluon exchange. Even in inclusive scattering factorization has
only been proved for a subset of the measureable hard processes \cite{fact}.

\begin{figure}[htb]
\begin{center}
\leavevmode
{\epsfxsize=13.5truecm \epsfbox{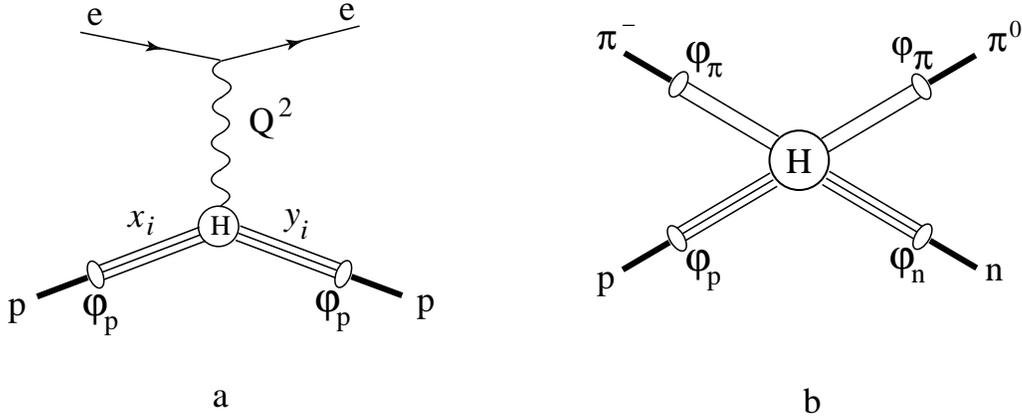}}
\end{center}
\caption[*]{a. Elastic $ep\to ep$ scattering at large $Q^2$ factorizes into a
product of proton distribution amplitudes $\varphi_p$ and a hard electron
scattering from the compact valence Fock state $\ket{uud}$. b. An analogous
factorization is illustrated for the large angle process $\pi^-p \to \pi^0n$.}
\label{fig1}
\end{figure}

The hard subprocess can occur coherently off several partons if the distance
between them is commensurate with the momentum transfer $Q$. Such (`higher
twist') processes are more strongly damped in momentum transfer than DIS
(\cf\ Eqs. (\ref{siginc}) and (\ref{sigexc})), since the partons must be
increasingly close as $Q$ grows. This is what happens in hard exclusive
processes, where factorization is also expected to apply \cite{brle}.
As an example, consider elastic electron-proton scattering, $ep\to
ep$ at large momentum transfer $Q$ (Fig. \ref{fig1}a). The amplitude for this
process has been shown to factorize into a product of a hard
scattering part $T_H$ and proton `distribution amplitudes' $\varphi_p$,
\beq
A(ep\to ep) = \int_0^1 \prod_{i=1}^3 dx_i dy_i \varphi_p(x_i,Q^2)\, T_H\,
\varphi_p(y_i,Q^2) \{1+\morder{1/Q^2}\}  \label{epep}
\eeq

The proton distribution amplitude is the valence part of the Fock expansion
(\ref{fock}), integrated over relative transverse momenta up to $Q$,
\beq
\varphi_p(x_i,Q^2) = \int^{k_{i\perp}^2 < Q^2} \prod_i d^2\vec k_{i\perp}
\Psi_{uud}(x_i,k_{i\perp})  \label{distamp}
\eeq
where the $x_i$ denote the longitudinal momentum fractions of the valence
$uud$ constituents. The integral over the relative transverse momenta
$k_{i\perp}$ implies that the transverse size of the valence state is
$r_\perp \simeq 1/Q$. The hard amplitude $T_H$ describes the subprocess
$e+(uud) \to e+(uud)$, which selects compact $\ket{uud}$ states.

The logarithmic $Q^2$ dependence of the proton distribution amplitude is given
by
\beq
\varphi_p(x_i,Q^2)=120 x_1 x_2 x_3 \delta(1-x_1-x_2-x_3) \sum_{n=0} \left[
\frac{\as(Q^2)}{\as(Q_0^2)} \right]^{\lambda_n} C_n P_n(x_i)
\label{qdep}
\eeq
The anomalous dimensions form an increasing series
\beq
\lambda_0= \frac{2}{27} < \lambda_1= \frac{20}{81} < \lambda_2= \frac{24}{81} <
 \ldots
\eeq
implying that each successive term in \eq{qdep} decreases faster with $Q^2$
than the previous one. The $P_n$ are Appell polynomials, $P_0=1$, 
$P_1=x_1-x_3$, $P_2= 1-3x_2,\ \ldots$ and the $C_n$ are constants which
characterize the proton wave function and have to be determined from
experiment. The $Q^2$ evolution of the pion distribution amplitude is given by
an expression similar to \eq{qdep}. The overall normalization of the
pion distribution amplitude is fixed by the decay constant $f_\pi$ measured in
$\pi \to \mu\nu$ decay.

Just as in the case of inclusive scattering, the relevance of
factorization for data on exclusive reactions must be demonstrated by showing
that the same (universal) distribution amplitudes $\varphi_h$ describe
several hard exclusive processes. For example, large angle $\pi^-p \to \pi^0 n$
scattering should be described by the diagram of Fig. \ref{fig1}b, which
involves the pion and proton distribution amplitudes and the $(q\bar
q)+(qqq)$ elastic subprocess.  Heavy meson decays like $B
\to \pi\pi$ can also be analyzed in the same formalism (assuming that the
momentum transfers involved are large enough).

Tests of factorization in exclusive reactions are quite difficult in
practice. From a theoretical point of view, the calculation of
multi-parton scattering amplitudes like those in Fig. \ref{fig1} are very
demanding even at the Born level, due to the large number of Feynman
diagrams. It is also difficult to estimate how high momentum
transfers are required in order to reach the scaling regime. Thus in Fig.
\ref{fig1}a the momentum transfer $Q$ from the electron is effectively
split among the three quarks of the proton. The less momentum a quark
carries, the less transfer it needs to scatter to a large angle. There is
an especially dangerous region where some of the valence quarks carry a
very small fraction $x$ of the proton momentum, in which case they can
fit into the proton wave function both before and after the hard
scattering, without receiving any momentum transfer. There has been much
discussion as to the importance of this `Feynman mechanism' \cite{isgll}. The
consensus appears to be that it is suppressed asymptotically \cite{bost} due
to the Sudakov effect \cite{suda}: The single quark carrying all the momentum
cannot be deflected to a large angle without gluon emission. At finite (and
relevant) energies, the importance of the Feynman mechanism is still not
settled -- and its significance may depend on the reaction.

An immediate consequence of factorization for exclusive reactions is the
`counting' or 'dimensional scaling' rule \cite{dsr}, which gives the power of
the squared momentum transfer $t$ by which any $2\to 2$ fixed angle
differential cross section is suppressed (up to logarithms),
\beq
\frac{d\sigma}{dt}(2\to 2) \propto \frac{f(t/s)}{t^{n-2}}  \label{scarul}
\eeq
where $n$ is the total number of elementary fields (quarks, gluons, photons)
that are involved in the scattering. This rule follows from simple
geometrical considerations. Elastic scattering between two elementary
fields (\eg, $qq\to qq$) involves no dimensionful quantities except $s$ and
$t$ and thus obeys \eq{scarul} with $n=4$ at fixed $t/s$. Each additional field
that is involved in the scattering must be within a transverse distance of
order $r_\perp \lsim 1/Q$ (with $Q^2 = -t$) to scatter coherently, and the
probability for that is of
\order{1/(Q^2 R^2)}, where $R \simeq 1$ fm is the average radius of the
hadron. This rule also explains why the dominant contribution to hard
scattering comes from the valence Fock states, which minimize the power $n$
in \eq{scarul}.

It is encouraging (although by no means conclusive) for factorization in
hard exlusive processes that the scaling rule (\ref{scarul}) is
approximately obeyed by the data for many reactions. Thus, $ep \to ep$
involves a minimum of $n=8$ fields, implying that the
proton form factor should scale as $F_p(Q^2) \propto 1/Q^4$, as assumed in
\eq{sigexc}. Data is available
\cite{pform} for $Q^2 \lsim 30\ \gev^2$ and is consistent with this behavior
for $Q^2 \gsim 5\ \gev^2$. At the higher values of $Q^2$ there are indications
of scaling violations that are consistent with the logarithmic evolution
predicted by \eq{epep}.

Tests of the dimensional scaling rules in exclusive reactions are
analogous to tests of Bjorken scaling in DIS, \ie, that the $Q^2$
dependence of the inclusive cross section is given by \eq{siginc}. In DIS,
the cross section as a function of $x$ then directly measures the structure
function $F(x)$. In exclusive reactions the situation is not as favorable.
The experimentally determined normalization of the proton form factor only
gives us one number, which is an average of the proton
distribution amplitude integrated over the momentum fractions $x_i$ carried
by the valence quarks. To make a quantitative prediction one must know both
the shape and the normalization of the (non-perturbative) distribution
amplitude. The good news is that the asymptotic form of the amplitude in the
$Q^2\to\infty$ limit is known, $\varphi_p^{AS} \propto x_1x_2x_3$ according to
\eq{qdep}. The non-asymptotic corrections are encoded in the moments $C_i$
which are measurable in principle. Considerable efforts have been made
to determine the pion and proton distribution amplitudes theoretically using
lattice calculations and QCD sum rules \cite{sumrules,rads}.

\begin{figure}[htb]
\begin{center}
\leavevmode
{\epsfxsize=13.5truecm \epsfbox{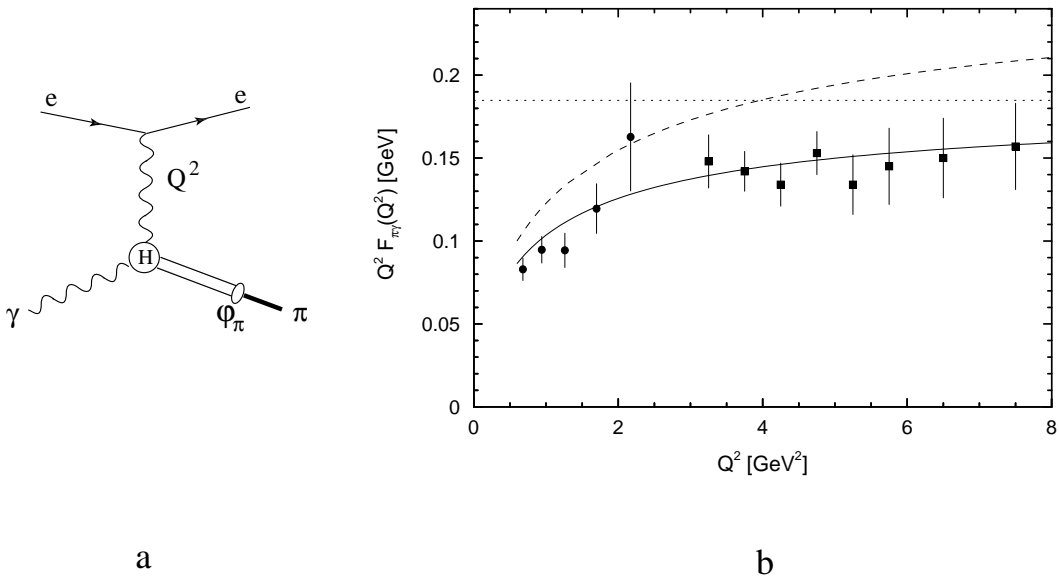}}
\end{center}
\caption[*]{a. The pion transition form factor $F_{\pi\gamma}$ is measured
by the process $e\gamma \to e\pi$, and factorizes at high $Q^2$ into a
product of the calculable hard subprocess $e\gamma \to e+(q\bar q)$ and the
pion distribution amplitude $\varphi_\pi$. b. Data \cite{pida}
compared with calculations based on a pion distribution amplitude
close to the asymptotic one (solid line) and one based on QCD sum rules
\cite{sumrules} (dashed line). The dotted line represents the asymptotic
result $\sqrt{2}f_\pi$. Figure from Kroll \etal\ in \cite{pitff}.}
\label{fig2}
\end{figure}

One of the simplest hard exclusive processes is the pion transition form
factor $F_{\pi\gamma}(Q)$, measured by the process $e\gamma \to e\pi$ at large
momentum transfer $Q$, \cf\ Fig. \ref{fig2}a. The existing data \cite{pida} in
the range
$1<Q^2<8\ \gev^2$ shown in Fig. \ref{fig2}b is well fit using a pion
distribution amplitude close to the asymptotic form $\varphi_{\pi}^{AS}
\propto x_1x_2$ \cite{pitff}. Considering that the absolute normalization in
the large $Q^2$ limit is fixed by the pion decay constant,
$F_{\pi\gamma}^{AS}=\sqrt{2}f_\pi/Q^2$, the agreement is very encouraging and
indicates that the factorization formalism applies even at moderate
values of $Q^2$. There is evidence, on the other hand, that the asymptotic
regime may be more distant in the case of the pion form factor measured by
$e\pi \to e\pi$ large angle scattering \cite{piff}.

There are many other processes that can and need to be
analyzed experimentally and theoretically in order to achieve a
comprehensive understanding of the phenomenology of hard exclusive
scattering. A particularly important process is virtual compton scattering
$\gamma^*p\to \gamma p$, which involves no hadrons except the proton and
offers the possibility of varying independently both the virtuality of the
photon and the momentum transfer to the proton \cite{shupe,krni,rads,ji}. Many
exclusive processes involve resonance production and thus require the
measurement of multiparticle final states. It seems clear that the
phenomenology of rare exclusive processes requires the capability of an \elf\ 
type accelerator, which combines sufficient energy with high luminosity in a
continuous electron beam.

\subsection{Scattering from Compact Subsystems}

Both in inclusive DIS and in hard exclusive processes a photon (or gluon)
scatters from a parton system ($q, g, q\bar q$ or $qqq$) with a transverse size
of \order{1/Q}, compatible with the photon wavelength. Intuitively, this is
required for the physics of the hard perturbative scattering to factorize from
the non-perturbative wave function, which determines the probability for such
compact systems. 

Fully inclusive scattering like DIS measures single parton distributions, with
no constraint on the size of the Fock state to which they belong. In exclusive
scattering the whole Fock state is required to be compact. There are also
intermediate (semi-inclusive) hard processes where the scattering occurs off
{\em multiparton subsystems} of the hadron, such as
$qq,\ gg,\ etc.$ The theoretical framework for such processes is still
incomplete, but the factorization of hard and soft physics seems plausible.
This would allow experimental measurements of the compact subsystems and of
the momentum fraction that they carry.

\begin{figure}[htb]
\begin{center}
\leavevmode
{\epsfxsize=13.5truecm \epsfbox{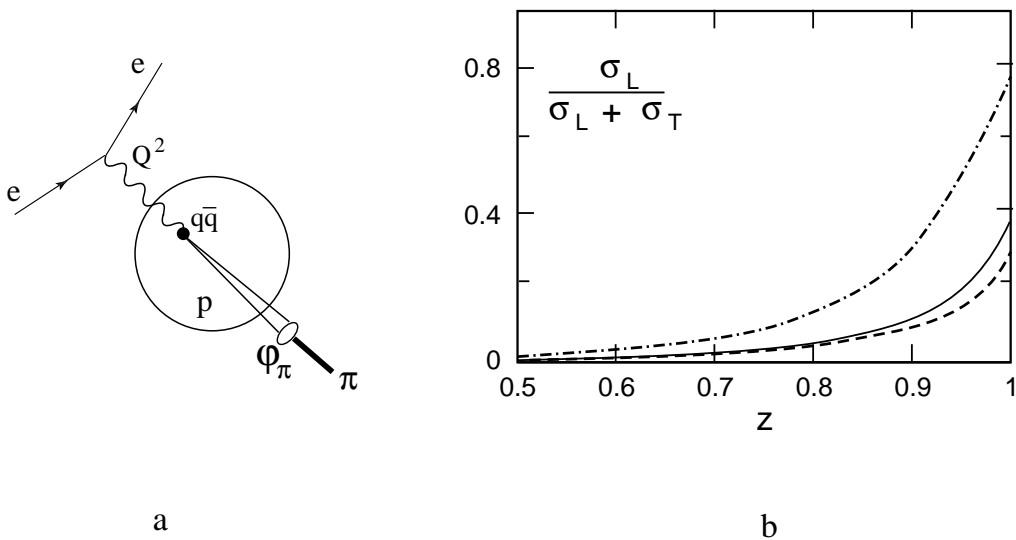}}
\end{center}
\caption[*]{a. Electron scattering off compact $\qpair$ pairs in the target
are selected by the semi-inclusive process $ep\to e\pi+X$ when the pion
carries a large fraction $z$ of the photon energy. b. Model
calculation \cite{bra} of the ratio $\sigma_L/(\sigma_L+\sigma_T)$, showing
how coherent scattering on $\qpair$ begins to dominate at large $z$. The
curves correspond to different choices of the pion distribution amplitude.}
\label{fig3}
\end{figure}

As an example, consider the semi-inclusive process $ep \to e\pi + X$
sketched in Fig. \ref{fig3}a, where the pion takes a fraction $z$ of
the photon energy $\nu$. In the limit $z\to 1$ the photon transfers all
its energy to the pion, which selects compact $\qpair$
configurations \cite{bb,bra}. Alternatively (and in fact equivalently), the
struck quark needs to combine with a very soft antiquark to form the pion --
such asymmetric configurations are short-lived and indistinguishable (by the
photon) from compact $q\bar q$ pairs \cite{bhmt}. Thus the cross
section can be expected to factorize in the $z\to 1$ limit as
\beq
\sigma= \hat\sigma(e+(\qpair) \to e+(\qpair))\,F_{\qpair/p}(x)\ |\varphi_\pi|^2
\label{piprod}
\eeq
where $F_{\qpair/p}(x)$ is the probability for finding the compact quark
pair in the target, and the pion distribution amplitude $\varphi_\pi$ is
the amplitude for the pair to transform into a physical pion.

Scattering off $q\bar q$ pairs (having integer spin) can be distinguished
from scattering off single (spin 1/2) quarks through the ratio
$\sigma_L/(\sigma_L+\sigma_T)$ of the longitudinally polarized to total
photon cross sections. As is well known, $\sigma_L=0$ (up to higher order QCD
corrections) for scattering from spin 1/2 quarks, whereas $\sigma_T=0$
for scattering on spin 0 diquarks. A calculation of the cross section ratio as
a function of $z$ based on the model orginally proposed in Ref. \cite{bb} is
shown in Fig. \ref{fig3}b. Experimental evidence for an analogous effect has
been seen in the reverse reaction $\pi N\to \mu^+\mu^- + X$, where the muon
pair takes a high fraction $x_F$ of the pion momentum \cite{e615}.

Systematic high statistics studies of semi-inclusive reactions for several
targets will allow measurements of both the diquark structure function
$F_{\qpair/p}(x)$ in \eq{piprod} and of nuclear effects on the pion
distribution amplitude $\varphi_\pi$, due to incomplete color transparency. A
precise theoretical formulation of scattering from subsystems is
called for and progress in this direction is being made \cite{rads,ji,dvcs}.

\section{Short Range Correlations in Nuclei}

The inclusive nuclear structure function is to a first approximation
given by the nucleon one, $F_{q/A}(x) \simeq A F_{q/N}(x)$ \cite{arn}.
Deviations of
\order{20 \ldots 30\ \%} are observed for small values of $x$ (`shadowing') and
for $x=0.5 \ldots 0.7$ (the `EMC effect'). When viewed in coordinate space, one
finds \cite{hova} that the quark `mobility distribution' is almost independent
(at the 2\% level) of $A$ up to light-cone distances (conjugate to $Q^2/2\nu$)
of order 2 fm, with shadowing setting in at larger distances. Since DIS (at
moderate $x$ and in coordinate space) is dominated by the most common Fock
states, this result shows that typical nucleon configurations are little
affected by the nuclear environment. The shadowing effect at large light-cone
distances reflects coherent scattering off several nucleons in the
nucleus.

In contrast to inclusive scattering, hard semi-inclusive and exclusive
scattering select rare parton configurations, where some or all of the
partons in the Fock state are at short relative transverse distance. Since
such configurations do not contribute to DIS at moderate values of $x$ their
$A$-dependence is essentially unknown. Clusters that carry more momentum than
single nucleons in the nucleus are of special interest, since they select
nuclear configurations where several nucleons are at short relative distance.
In the parlance of nuclear physics, these represent highly excited states of
the nucleus (with excitation energies in the GeV region) about which we know
very little at present. An electron beam of high intensity and resolution is
essential for mapping out such dense clusters.

In DIS on nuclei, the fraction $x=Q^2/2m_p\nu$ of the target momentum carried
by the struck quark has the range $0 \leq x \leq A$. Data at $x \gsim
1$ exists and is difficult to explain by standard Fermi
motion \cite{bcdms,slac,nutev}. Models based on short-range correlations
between nucleons \cite{frastri} and on multi-quark effects \cite{bags} can fit
the data, but considerably more experimental and theoretical effort
will be needed to clarify the physics of this `cumulative' region of nuclei.

Novel cumulative effects are observed also in nuclear
fragmentation into hadrons \cite{frastri,stav,geag}. The hadron ($p,\ \pi,\ K$)
momentum distributions extend beyond $x_F=1$, \ie, their momentum must have
been transferred from several nucleons. The fragmentation is only weakly
dependent on the nature of the projectile or its energy, indicating that it
measures features intrinsic to the nuclear wave function. In these processes the
projectile scattering is soft, but there is evidence \cite{boya} that the
average transverse momentum of the produced hadrons increases with $x_F$,
reaching $\langle p_\perp^2 \rangle = 2\ \gev^2$ at $x_F=4$ for protons. The
cumulative momentum transfers thus appear to originate in a transversally
compact region of the nucleus.

Cumulative nuclear effects have furthermore been observed in subthreshold
production of antiprotons and kaons \cite{subth}. The minimal projectile
energy required for the process $pp\to \bar p+X$ on free protons
at rest is 6.6 GeV. The kinematic limit for $pA\to \bar p+X$ on a heavy
nucleus at rest is only $3m_N\simeq 2.8$ GeV. This reaction has been observed
for $A$ = $^{63}$Cu down to $E_{lab}\simeq3$ GeV, very close to kinematic
threshold. Scattering on a single nucleon in the nucleus would at this energy
require a Fermi momentum of \order{800} MeV. While the $pA$ data can be fit
assuming such high Fermi momenta, this assumption leads to an underestimate of
subthreshold production in $AA$ collisions by about three orders of magnitude
\cite{fermom}. 

It is possible that the subthreshold production of $K$ and $\bar p$ on nuclei
involves the same compact multiparton clusters that are responsible for
scattering with $x>1$ and $x_F>1$, although this is far from clear at present.
A study of subthreshold production using lepton beams could be quite
informative, since the locality of the reaction can be tuned through the
virtuality of the exchanged photon. A further possibility to pin
down the reaction mechanism is provided by subthreshold production of charm.
ELFE will in fact be working close to charm threshold, allowing for many
interesting phenomena. A discussion of charm physics near threshold can be
found in Ref. \cite{charm}.

\section{Conclusions}

There are (at least) three central physics areas which require an
accelerator with the capabilities of ELFE as given in Table 1:
\begin{itemize}
\item[$\bullet$] The determination of hadron and nuclear wave functions.
\item[$\bullet$] Specifically nuclear effects: Color transparency \cite{ct},
cumulative phenomena \cite{bcdms} -- \cite{fermom}.
\item[$\bullet$] Charm(onium) production near threshold \cite{charm}.
\end{itemize}

In addition to these core topics there are a number of areas where ELFE
can improve on presently available data, such as 
\begin{itemize}
\item[--] The nucleon structure function for $0.7\lsim x \lsim 1$,
\item[--] Higher twist corrections of the form $c(x)/Q^2$,
\item[--] $R=\sigma_L/\sigma_T$,
\item[--] The gluon structure function,
\item[--] Polarized structure functions.
\end{itemize}
Significant advances in these areas are, however, expected from other
experiments before ELFE starts operating.

Finally, we should keep in mind that the whole area of `confinement' physics
is very important but at present poorly understood in QCD. It includes open
questions like the influence of the QCD vacuum on scattering processes
\cite{nacht} and the foundations of the non-relativistic quark model (see,
\eg, \cite{diak,eps}). It is difficult to assess today what the progress will
be in this field. Nevertheless, it seems clear that systematic
measurements of non-perturbative wave functions as discussed above will form an
essential part of any serious effort to understand the hadron spectrum.   

{\bf Acknowledgements.} I would like to thank the organizers of this meeting
for arranging a very interesting cross-disciplinary discussion of the research
possibilities at a future DESY linear collider. My understanding of the
material discussed above stems mainly from a long and fruitful collaboration
with Stan Brodsky. I am also grateful for helpful discussions with V. M. Braun
and M. Strikman.


\begin{thebibliography}{99}

\bibitem{wiik}
B. H. Wiik, {\em Future Electron Accelerators and Free Electron Lasers}, Talk
given at this meeting.

\bibitem{frois}
B. Frois, {\em Nuclear Physics -- Experimental Aspects}, Talk
given at this meeting; A. Tkatchenko, {\em Machine Project for ELFE at DESY},
Talk given at the Second ELFE Workshop, St. Malo, 22-27 September 1996.

\bibitem{charm} P. Hoyer, {\em Charmonium Production at ELFE Energies,} talk at
the 2nd ELFE Workshop, Saint Malo, France, Sept. 1996 (to be published).

\bibitem{elphys}
B. Frois and B. Pire, Invited talk at 8th International Nuclear
Physics Conference (INPC 95), Beijing, P.R. China, 21-26 Aug 1995,
hep-ph/9512221; J. Arvieux and B. Pire, Prog. Part. Nucl. Phys. {\bf 35} (1995)
299. 

\bibitem{bro} S. J. Brodsky, Talk presented at Orbis Scientiae (Miami Beach
1996), SLAC-PUB-7152, hep-ph/9604391.

\bibitem{ct}  N. N. Nikolaev and B. G. Zakharov, Talk at the INPC Conference,
Beijing, China, August 1995, KFA-IKP(TH)-1995-21, nucl-th/9509036; P. Hoyer,
Talk at Workshop on Deep Inelastic Scattering and QCD (DIS 95), Paris,
France, April 1995, proceedings Paris DIS 1995:127, hep-ph/9510394.

\bibitem{fact} J. C. Collins, D. E. Soper and G. Sterman, in
{\em Perturbative QCD}, ed. A.H. Mueller (World Scientific, 1989); 
G. Bodwin, \PR{D31}{1985}{2616} and {\bf D34} (1986) 3932 (E); 
J. Qiu and G. Sterman, \NP{B353}{1991}{105} and {\bf B353} (1991) 137.

\bibitem{brle} For a review, see S. J. Brodsky and G. P. Lepage in {\em
Perturbative QCD}, edited by A. H. Mueller (World Scientific, Singapore,1989).

\bibitem{isgll} N. Isgur and C. H. Llewellyn-Smith, \NP{B317}{1989}{526};
A. V. Radyushkin, \NP{A532}{1991}{141c}.

\bibitem{bost} J. Botts and G. Sterman, \NP{B325}{1989}{62}.

\bibitem{suda} V. V. Sudakov, Sov. Phys. JETP {\bf 3} (1956) 65. 

\bibitem{dsr} S. J. Brodsky and G. R. Farrar, \PRL{31}{1973}{1153}; V. A.
Matveev, R. M. Muradyan and A. V. Tavkhelidze, Lett. Nuovo Cimento {\bf 7}
(1973) 719.

\bibitem{pform} A. F. Sill \etal, \PR{D28}{1993}{860}.

\bibitem{sumrules} V. L. Chernyak and A. R. Zhitnitsky,
\PRe{112}{1984}{173}; S. V Mikhailov and A. V. Radyushkin,
\PR{D45}{1992}{1754}; A. V. Radyushkin and R. Ruskov, \PL{B374}{1996}{173}.

\bibitem{rads} A. V. Radyushkin, Talk at Workshop on Virtual Compton
Scattering, Clermont-Ferrand, France, June 1996, JLAB-THY-96-06,
hep-ph/9609387.

\bibitem{pida} CELLO Collaboration, H.-J. Behrend \etal, \ZP{C49}{1991}{401};
CLEO Collaboration, V. Savinov \etal, proceedings of the PHOTON '95 Workshop
Sheffield, England, April 1995, hep-ex/9507005.

\bibitem{pitff} A. V. Radyushkin and R. Ruskov, CEBAF-TH-95-18,
hep-ph/9603408; P. Kroll and M. Raulfs, \PL{B387}{1996}{848},
hep-ph/9605264.

\bibitem{piff} R. Jakob and P. Kroll, \PL{B315}{1993}{463}; {\bf B319} (1993)
545 (E).

\bibitem{shupe} M. A. Shupe \etal, \PR{D19}{1979}{1929}.

\bibitem{krni} A. S. Kronfeld and B. Ni\v zi\'c, \PR{D44}{1991}{3445};
G. R. Farrar, K. Huleihel and H. Zhang, \NP{B349}{1991}{655}.

\bibitem{ji} X. Ji, MIT-CTP-2568, hep-ph/9609381.

\bibitem{bb} E. L. Berger and S. J. Brodsky, \PRL{42}{1979}{940}; A.
Brandenburg, S. J. Brodsky V. V. Khoze and D. Muller,
\PRL{73}{1994}{939}, hep-ph/9403361; K. J. Eskola, P. Hoyer, M. V\"anttinen
and R. Vogt, \PL{B333}{1994}{526}, hep-ph/9404322.

\bibitem{bra} A. Brandenburg, V. V. Khoze and D. Muller,
\PL{B347}{1995}{413}, hep-ph/9410327.

\bibitem{bhmt} S. J. Brodsky, P. Hoyer, A. H. Mueller and W.-K. Tang,
\NP{B369}{1992}{519}.

\bibitem{e615} E615 Collaboration, J. S. Conway \etal, \PR{D39}{1989}{92}.

\bibitem{dvcs} X. Ji, MIT-CTP-2517, hep-ph/9603249;  A. V. Radyushkin,
CEBAF-TH-96-06, hep-ph/9605431.

\bibitem{arn} M. Arneodo, \PRe{240}{1994}{301}.

\bibitem{hova} P. Hoyer and M. V\"anttinen, NORDITA-96-20-P, hep-ph/9604305. 

\bibitem{bcdms} BCDMS Collaboration, A. C. Benvenuti \etal,
\ZP{C63}{1994}{29}.

\bibitem{slac} J. Arrington \etal, \PR{C53}{1996}{2248}.

\bibitem{nutev} CCFR/NuTeV Collaboration, M. Vakili \etal, 
in {\it Proceedings of the Division of Particles and Fields meeting, 1996
(DPF96)}, Minneapolis, USA, August, 1996.

\bibitem{frastri} L. L. Frankfurt and M. Strikman, \PRe{160}{1988}{325}; L. L.
Frankfurt, M. I. Strikman, D. B. Day and M. Sargsyan, \PR{C48}{1993}{2451}.

\bibitem{bags} A. V. Efremov, Sov. J. Part. Nucl. {\bf 13} (1982) 254; L.
Kaptari and A. Umnikov, JINR Rapid Commun. {\bf 32} (1988) 17; S. Gupta and R.
M. Godbole, \PL{228B}{1989}{129}.

\bibitem{stav} V. S. Stavinskii, Sov. J. Part. Nucl. {\bf 10} (1979) 373.

\bibitem{geag} J. V. Geagea \etal, \PRL{45}{1993}{1980}; A. Gillitzer \etal,
\ZP{A354}{1996}{3}.

\bibitem{boya} S. V. Boyarinov \etal, Sov. J. Nucl. Phys. {\bf 46} (1987) 871.

\bibitem{subth} J. B. Carroll \etal, \PRL{62}{1989}{1829}; A. Shor \etal,
\PRL{63}{1989}{2192}; A. Schr\"oter \etal, \ZP{A350}{1994}{101}.

\bibitem{fermom} A. Shor, V. Perez-Mendez and K. Ganezer,
\NP{A514}{1990}{717}.

\bibitem{nacht} O. Nachtmann, Johns Hopkins Workshop 1994:143-172, 
hep-ph/9411345.

\bibitem{diak} D. Diakonov, Talk at International School of Nuclear
Physics, Erice, Italy, Sept. 1995, nucl-th/9603023.

\bibitem{eps} P. Hoyer, NORDITA-96/63 P, hep-ph/9610270.


\end{thebibliography}
\end{document}